\documentclass[12pt,psfig]{article}
\usepackage{epsf,graphicx}

\newcommand{\beq}{\begin{equation}}
\newcommand{\eeq}{\end{equation}}
\newcommand{\be}{\begin{eqnarray}}
\newcommand{\ee}{\end{eqnarray}}

\def\nue{{\nu_e}}

\newcommand{\dm}{\mbox{$\Delta{m}^{2}$~}}

\begin{document}
\title{\bf Neutrino decay confronts the SNO data} 
\author
{Abhijit Bandyopadhyay\thanks{e-mail: abhi@theory.saha.ernet.in},   
 Sandhya Choubey\thanks{email: sandhya@hep.phys.soton.ac.uk}, 
 Srubabati Goswami\thanks{e-mail: sruba@mri.ernet.in}
\\
\\
$^*${\it Theory Group, Saha Institute of Nuclear Physics},\\
{\it 1/AF, Bidhannagar,
Calcutta 700 064, INDIA}\\
$^\dagger${\it Department of Physics and Astronomy, University of Southampton}, \\
{\it Highfield, Southampton S017 1BJ, UK}\\
$^\ddagger${\it Harish-Chandra Research Institute},\\{\it Chhatnag Road, Jhusi, 
Allahabad  211 019, INDIA}}
\maketitle

\begin{abstract}

We investigate the  status of the neutrino decay solution to 
the solar neutrino problem in the context of the recent results
from Sudbury Neutrino Observatory (SNO). 
We present the results of  global $\chi^2$-analysis for both two and 
three generation cases with one of the mass states
being allowed to decay and include the effect of both decay and mixing.
We find that the  
Large Mixing Angle (LMA) region which is the currently favoured 
solution of the solar neutrino problem is affected significantly 
by decay.  
We present the allowed areas in the $\Delta m^2-\tan^2\theta$ 
plane for different allowed values of $\alpha$ and examine  
how these areas change 
with the inclusion of 
decay. 
We obtain bounds on the 
decay constant $\alpha$ in this region which implies
a rest frame life time $\tau_0/m_{2} > 8.7 \times 10^{-5}$ 
sec/eV for the unstable neutrino state. 
We conclude that the arrival of the neutral current results from 
SNO further disfavors the neutrino decay solution to the 
solar neutrino problem leaving a very small window for the 
decay constant $\alpha$ which could still be allowed.


\end{abstract}

\newpage

\section{Introduction}


The comparison of the SNO charged current (CC) measurments with their 
neutral current (NC) rate establishes oscillations to active neutrino 
flavor at the $5.3\sigma$ level \cite{Ahmad:2002jz}. The inclusion of 
Super-Kamiokande (SK) electron scattering (ES) rate \cite{Fukuda:2002pe}
with SNO confirms transitions 
to active flavors at more than $5.5\sigma$ level. In other words SNO 
rules out the possibility of having 
a purely sterile component in the solar neutrino 
beam at the $5.5\sigma$ level. 
However  presence of sterile components in the beam  
is not entirely ruled out. Transition to {\it mixed} 
states\footnote{States which are a mixture of active as well as 
sterile components.} is still allowed with  $<32\%(65\%)$ sterile 
mixture at the $1\sigma(2\sigma)$ level 
\cite{Bandyopadhyay:2002xj,Bahcall:2002zh,Maltoni:2002ni}.
One of the scenarios in which one can have a
final sterile state is neutrino decay.
The aim of this paper is to investigate the status
of the neutrino decay solution to the solar neutrino problem with the 
incorporation of the SNO data. 
We deal with both two generation and three generation cases. 
For the latter we consider  
three flavor mixing between
$\nu_e$, $\nu_\mu$ and $\nu_\tau$ with 
mass eigenstates $\nu_1$, $\nu_2$ and $\nu_3$ 
with the assumed mass hierarchy as $m^2_1<m^2_2<m^2_3$.
The lightest neutrino mass state is assumed to have lifetime much
 greater than the 
Sun-Earth transit time and hence can be taken as stable.
Whereas the  heavier mass states $\nu_2$ and $\nu_3$
may be unstable.
But the CHOOZ data \cite{chooz} constrains the mixing matrix element 
U$_{e3}$ to a very small value which implies  
very small mixture of the $\nu_3$ state to the $\nu_e$
state produced in Sun. So the instability of
the $\nu_3$ state has hardly any effect on solar 
neutrino survival probability, and to study the 
effect of decay on solar neutrinos for simplicity we can consider only
the second mass state $\nu_2$ to be unstable. 

There are two kinds of models of non-radiative decays of neutrinos 
\begin{itemize}
\item If neutrinos are Dirac particles 
one has the decay channel $\nu_2 \rightarrow \bar{\nu}_{1R} + \phi$, 
where ${\bar{\nu}}_{1R}$ is a right handed singlet and
 $\phi$ is an iso-singlet scaler \cite{Acker:1990cq}.
 Thus all the final state particles are sterile. 
\item If neutrinos are Majorana particles, 
the decay mode is $\nu_2 \rightarrow \bar{\nu}_1 + J$,
 where $\bar{\nu}_1$
interacts as a $\bar{\nu}_e$ with a probability 
$|U_{e1}|^2$ and 
J is a Majoron \cite{Acker:1992eh}.
\end{itemize}

In both scenarios the rest frame lifetime of $\nu_{2}$ is given by 
\cite{Acker:sz}
\begin{equation}
\tau_{0} = \frac{16 \pi}{g^2} \frac{m_{2} (1 + m_{1}/m_{2})^{-2}}{\dm}
\label{tau0}
\end{equation}
where $g$ is the coupling constant and $\Delta m^2 (= m_2^2 - m_1^2$) is 
the mass squared difference between the states involved in the decay 
process. 
If a neutrino of energy E decays while traversing a distance $L$ then the 
decay term exp(-$\alpha L/E$) gives the fraction of neutrinos that decay. 
$\alpha$ is the decay constant and is related to $\tau_{0}$ as 
$\alpha = m_{2}/\tau_{0}$. 

For the Sun-Earth distance of $1.5 \times 10^{11}$ m,
and for a typical neutrino energy of 
10 MeV  one starts getting appreciable decay for  $\alpha \sim 10^{-12}$ 
eV$^2$. For lower  values of $\alpha$ the exp(-$\alpha L/E)$ term goes to 
1 signifying no decay while for $\alpha$ $>$ $10^{-10}$ eV$^2$ the 
exponential
term goes to zero signifying complete decay of the unstable neutrinos.  
Assuming
$m_{2} >> m_{1}$  the equation (\ref{tau0}) can be written as
\begin{equation}
g^2 \dm \sim 16 \pi \alpha 
\label{galpha}
\end{equation} 
If we now incorporate the bound $g^2 < 4.5 \times 10^{-5}$ as obtained
from K decay modes \cite{barger}
we get the bound $\Delta m^2 > 10^{6} \alpha$. This gives the  
range $\Delta m^2 \stackrel{>}{\sim} 10^{-6} $ eV$^2$ for which 
we can have appreciable decay.

Neutrino decay in the context of the solar neutrino problem has been 
considered earlier in \cite{Acker:1990cq,Acker:1992eh,Raghavan:1987uh} 
and more recently  
in  \cite{Choubey:2000an,Bandyopadhyay:2001ct,chou}. 
After the declaration of the SNO CC data
last year \cite{sno},  
bounds on the decay constant $\alpha$ have been obtained from the 
data on total rates \cite{Joshipura:2002fb} as well as from SK spectrum data 
\cite{Beacom:2002cb}. However with the declaration of the recent SNO 
neutral current results \cite{Ahmad:2002jz}
we expect to have a 
better handle on the sterile admixture in the solar neutrino 
beam and hence tighter constraints on the decay constant $\alpha$. 
The statistical analysis for the two-generation oscillation solution 
for stable neutrinos with the global solar neutrino data 
including the recent results of the SNO experiment can be found in 
\cite{Bandyopadhyay:2002xj,Fogli:2002pt,Bahcall:2002hv}.
In this paper we consider the possibility for unstable neutrino 
states and do a combined global satistical analysis of the solar 
neutrino data including the total rates from the Cl  
and Ga experiments \cite{globalsolar}, the zenith-angle recoil energy 
spectrum from SK \cite{Fukuda:2002pe} and the 
combined (CC+ES+NC) day-night energy spectrum from SNO \cite{Ahmad:2002jz}.  
We do our analysis for both two and three generation scenarios.
We include the results of the CHOOZ experiment \cite{chooz}
in our three generation analysis.

In section 2 we briefly present the survival probability for the 
solar neutrinos with one of the components unstable. In section 3 
we first find the constraints on $\alpha$ coming from 
the global solar neutrino data 
within a two-generation framework. 
We argue that decay of solar neutrinos 
is largely disfavored as a result of the 
inclusion of the SNO day-night spectrum. 
We next extend our analysis 
to include the third neutrino flavor and present bounds on $\alpha$, 
$\Delta m^2_{21}$ and $\tan^2\theta_{21}$ for different allowed 
values of $\theta_{13}$, from a $\chi^2$ analysis which includes both the 
global solar neutrino data as well as the CHOOZ data. 
We end with 
conclusions in section 4. 

\section{Formalism}

The three-generation 
mixing matrix relating the mass and flavour eigenstates 
are given as 
\be
U & = &
R_{23}R_{13}R_{12} \nonumber\\ & = &\pmatrix {c_{13}c_{12} &
s_{12}c_{13} & s_{13} \cr -s_{12}c_{23} - s_{23} s_{13} c_{12} &
c_{23} c_{12} - s_{23} s_{13} s_{12} & s_{23} c_{13} \cr s_{23}
s_{12} - s_{13} c_{23} c_{12} & -s_{23} c_{12} - s_{13} s_{12}
c_{23} & c_{23} c_{13} \cr} \label{mix} 
\ee
where we neglect the
CP violation phases. 

Allowing for the possibility of decay of the second mass eigenstate
the probability of getting a neutrino of flavour $f$ starting from an 
initial electron neutrino flavour is   
\footnote{For a rigorous derivation of the probability for the decay 
plus oscillation scenario see \cite{Lindner:2001fx}.}. 
\be
P_{ef} &=& {a_{e1}^\odot}^2|A_{1f}^\oplus|^2 + 
{a_{e2}^\odot}^2|A_{2f}|^2e^{-\alpha (L-R_{\odot})/E}
+ {a_{e3}^\odot}^2|A_{3f}^\oplus|^2 
\nonumber \\
&+& 2 a_{e1}^\odot a_{e2}^\odot e^{-\alpha (L-R_{\odot})/2E}
Re[A_{1f}^\oplus A_{2f}^{\oplus *}
e^{i(E_{2} - E_{1})(L - R_{\odot})} e^{i(\phi_{2,\odot} - \phi_{1,\odot})}]  
\label{pr}
\ee
where $E_i$ is the energy of the mass eigenstate $i$, $E$ is the 
energy of the neutrino beam, $\alpha$ is the decay constant, $L$ is the 
distance from the center of the Sun, $R_\odot$ is the radius of the Sun, 
$\phi_{i,\odot}$ are the phase inside the Sun and 
$a_{ei}^\odot$ is the amplitude of an electron state to be in the 
mass eigenstate $\nu_i$ at the surface of the Sun \cite{Bandyopadhyay:2001fb}  
\beq
{a_{ei}^{\odot}}^2 = \sum_{j=1,2,3} X_{ij} {{U^\odot_{je}}}^2
\eeq
where $X_{ij}$ denotes the non-adiabatic jump probability between
the $i^{th}$ and $j^{th}$ states inside the Sun 
and $U_{je}^\odot$ denotes the
mixing matrix element between the flavour state $\nu_e$ and the
mass eigenstate $\nu_j$ in Sun. 
$A_{if}^\oplus$ denote the $\nu_i  \rightarrow \nu_f$ 
transition amplitudes inside the Earth.
We evaluate these amplitudes numerically 
by assuming the Earth to consist of 
two constant density slabs \cite{Bandyopadhyay:2001fb}.
It can be shown that the square bracketed term
containing the phases averages out to 
zero in the range of \dm in which we are interested  \cite{Dighe:1999id}. 
For further details of the calculation of the survival probability 
we refer to \cite{Bandyopadhyay:2001fb}.

In the two generation limit 
the $\nue$ survival probability including Earth matter effects and decay 
can be expressed as \cite{Bandyopadhyay:2001ct} 
\begin{equation}
P_{ee} = P_{ee}^{\rm day} + \frac{(\sin^2 \theta - P_{2e})(P_{ee}^{\rm day}
+e^{-2 \alpha (L - R_{\odot})/E} (P_{ee}^{\rm day} -1))}{\cos^2 \theta - 
\sin^2 \theta e^{- 2 \alpha (L - R_{\odot})/E}}
\label{dn}
\end{equation}
where $P_{2e}$ is the transition probability of $\nu_2 \rightarrow \nu_e$  
at the detector while $P_{ee}^{\rm day}$ is 
the day-time survival probability (without Earth matter effects) 
for $\nue$ and is given by \cite{Bandyopadhyay:2001ct}
\begin{eqnarray}
P_{ee}^{\rm day} & = & \cos^2 \theta [ P_J \sin^2 \theta_M + (1 - P_J) \cos^2 \theta_M]
		\nonumber \\
       &  & + \sin^2 \theta [ (1 -P_J) \sin^2 \theta_M + \cos^2 \theta_M P_J]
	      e^{-2 \alpha (L - R_{\odot})/E}
\label{peesimple}
\end{eqnarray}
where
$P_{J}$ is the non-adiabatic level jumping probability 
between the two mass eigenstates for which we use the standard expression 
from \cite{Petcov:1987zj} and 
$\theta_M$ is the matter mixing angle given by
\beq
\tan 2\theta_{M} =  \frac{\Delta m^2\sin 2\theta}{\Delta m^2\cos2\theta - 
2\sqrt{2}G_{F}n_{e}E}.
\label{thetam}
\eeq
$n_{e}$ being the ambient electron density, $E$ the
neutrino energy, and \dm (= ${m_{2}^2 - m_{1}^2}$) the mass squared
difference in vacuum.

From eq.(\ref{peesimple}) we note that the
decay term appears with a $\sin^2\theta$ and is therefore 
appreciable only for large enough $\theta$. Thus we expect
the effect of decay to be maximum in the LMA region. 
This can also be understood as follows. 
The $\nu_e$ are produced mostly as $\nu_2$ in the solar core. 
In the LMA region the neutrinos move adiabatically through the Sun and 
emerge as $\nu_2$ which eventually decays. For the 
SMA region on the other hand $P_J$ is non-zero and $\nu_e$
produced as $\nu_2$ cross over to $\nu_1$ at the resonance and come out 
as a $\nu_1$ from the solar surface. 
Since  $\nu_1$ is stable, decay does not affect this region.

\section{Analysis of data and results} 

\subsection{Bounds from two-generation analysis}

First we use the standard $\chi^2$ minimisation procedure 
and determine the $\chi^2_{min}$ and the 
best-fit values of the oscillation and decay parameters, 
in a two flavor mixing scenario. 
Details of our statistical analysis procedure, including the 
definition of the $\chi^2$ and the correlated error 
matrix are given in Appendix A.
We incorporate the total 
rate in Cl and Ga, the zenith-angle spectrum data of SK and the 
SNO day-night spectrum.
We find that the global best-fit{\footnote{The 
bounds that we give here and 
subsequently apply to the decay model 
where both final states are sterile. In the Majoron decay model 
the final state $\bar{\nue}$ can interact  
in the SK/SNO detectors. However there is an energy degradation
and the best-fit values do not change significantly \cite{Choubey:2000an}. 
An interesting possibility 
where the absolute mass scales of the two neutrino states are 
approximately degenerate and hence the daughter $\bar\nu_e$ is not degraded 
in energy is considered recently in \cite{Beacom:2002cb}.}} 
comes in the LMA region 
with the decay constant $\alpha=0$ and 
$\tan^2\theta= 0.41,  \Delta m^2= 6.06 \times 10^{-5} 
$eV$^2$ \cite{proc}.

In figure 1 we plot the 
$\Delta \chi^2$ ($=\chi^2 - \chi^2_{min}$) 
for the two generation neutrino decay scenario, 
against the decay constant $\alpha$, keeping  
$\Delta m^2$ and $\tan^2\theta$ free. 
The figure shows that the
fit becomes worse with increasing value of the decay constant.  
We remind the reader that neutrino decay is important only for 
LMA and for each value of 
$\alpha$ on this curve the minimum $\chi^2$ comes with 
$\Delta m^2$ and $\tan^2\theta$ in the LMA region. 
The inclusion of the SNO data in the global solar neutrino 
analysis improves the fit in the LMA region for stable 
neutrinos\footnote{See 
\cite{Bandyopadhyay:2002xj,Fogli:2002pt,Bahcall:2002hv,proc} 
for recent global analyses and \cite{Bandyopadhyay:2001aa,sno2g} for 
earlier ones.}.
Hence for small 
values of $\alpha$ where neutrinos can be taken as almost stable, 
the inclusion of the SNO data gives a better 
fit. However as $\alpha$ increases the neutrino decay leads to more 
sterile components in the final state -- which is disfavoured by the 
SNO/SK combination and the 
fit worsens with SNO included. 
As is seen from figure 1 the global solar neutrino data 
put an upper bound on the decay constant -- 
$\alpha < 7.55 \times 10^{-12} $eV$^2$ at 99\% C.L., when $\Delta m^2$ 
and $\theta$ are allowed to take on any value. Before the declaration of the 
SNO results the bound on $\alpha$ from combined analysis of 
Cl, Ga and SK data was $\alpha < 3.5\times 10^{-11}$ eV$^2$ 
\cite{Bandyopadhyay:2001ct}. 
Thus the inclusion 
of the recent SNO results have further tightened the noose on the 
fraction of neutrinos decaying on their way from the Sun to Earth and 
neutrino decay is now barely allowed.

In figure 2 we plot the allowed areas in the 
$\Delta m^2 - \tan^2\theta$ plane for different allowed values 
of the decay constant $\alpha$ for a two flavor scenario and with global 
data including SNO.
The figure shows that the allowed area in the LMA region is 
reduced as $\alpha$ increases. 
As $\alpha$ increases the 
spectral distortion increases and the effect is more for 
higher(lower) values of $\tan^2\theta$($\Delta m^2$) 
\cite{Bandyopadhyay:2001ct}.
Since both the SK as well as the SNO spectra are consistent with 
no energy distortion,
these regions get disallowed with increasing $\alpha$. 
Spectral distortion due to decay is less for the high $\Delta m^2$. 
However these values for the $\Delta m^2$ are now disfavored at more 
than $3\sigma$ by SNO since the values for the $^8B$ flux required to 
explain the suppression in SK and SNO CC measurements are much 
lower than those consistent with the SNO NC observation.
In addition the low
energy neutrinos relevant for Ga experiment decay more  
making the fit to the total rates worse. 

As discussed earlier, for small values of mixing angles
the fraction of the decaying component in the $\nu_e$ beam 
($\sim \sin^2\theta$) being very small decay 
does not have much effect.
Hence no new feature is introduced in the SMA region
because of decay and it remains disallowed by the 
global data. 
On the other hand,
in the LOW region though mixing is large, due to small 
$\Delta m^2$ any appreciable decay over the Sun-Earth 
distance is not obtained, if the bound on the 
coupling constant from K-decay \cite{barger} is to be 
accounted for. Therefore, 
LOW region in the parameter space remains unaffected by the 
unstable $\nu_2$ state, and we do not show it explicitly in figure 2. 

\subsection{Bounds from three-generation analysis}

Next we investigate the impact of the unstable $\nu_2$ 
mass state on the three flavor oscillation scenario with
$\Delta m_{21}^2 = \Delta m_{\odot}^2 $ and 
$\Delta m_{31}^2 = \Delta m_{CHOOZ}^2 \simeq 
\Delta m_{atm}^2 = \Delta m_{32}^2. $ 
For the three generation case the $\chi^2$ is defined as 
\beq
\chi^2 = \chi^2_{solar} + \chi^2_{chooz}
\eeq
The $\chi^2_{solar}$ is defined in Appendix A while for 
the expression of $\chi^2_{chooz}$ 
we refer the reader to \cite{Bandyopadhyay:2001fb}. 
In figure 3 we plot the $\Delta \chi^2(=\chi^2-\chi^2_{min})$ vs $\alpha$ 
for various fixed 
values of $\theta_{13}$ allowing the other 
parameters to take arbitrary 
values.  $\Delta m_{31}^2$  is varied within the range
[1.5 - 6.0]$\times$ 10$^{-3}$ eV$^2$ as obtained
from a combined analysis of atmospheric and CHOOZ data
\cite{Gonzalez-Garcia:2000sq,Fogli:2001xt}\footnote{The range 
of $\Delta m_{31}^2$ allowed at 99\%C.L.  
from the combined analysis of final SK+K2K data is 
[2.3 - 3.1]$\times$ 10$^{-3}$ eV$^2$ \cite{Fogli:2002pb}. However 
the values of $\Delta m_{31}^2$ that we use are still allowed at 
the $3\sigma$ level.}.
This figure indicates the allowed range of $\alpha$
for a fixed $\theta_{13}$. 
As $\theta_{13}$ increases the 
contribution from $\chi^2_{chooz}$ increases shifting the plots 
upwards. 
Consequently the curve corresponding to higher $\tan^2\theta_{13}$ 
crosses the 99\% C.L. limit at a lower $\alpha$ and a more stringent
bound is obtained as compared to smaller $\tan^2\theta_{13}$
cases.  

In figure 4 we plot the allowed areas in the 
$\tan^2\theta_{12}-\Delta m_{21}^2$ plane
for
different sets of values of $\tan^2\theta_{13}$ and
$\Delta m_{31}^2$ at different values of $\alpha$.
For fixed values of $\tan^2\theta_{13}$  and $\Delta m^2_{31}$
the allowed area decreases with increasing $\alpha$
because of increased spectral distortion and transition of the solar 
neutrinos to sterile components. 
At any given $\alpha$ 
the allowed area shrinks with the increase of 
$\tan^2\theta_{13}$ and $\Delta m_{31}^2$.
This is again an effect of the CHOOZ data.

\section{Conclusions}

In this paper we have done a global $\chi^2$ analysis of the   
solar neutrino data assuming neutrino decay. We incorporate 
the  full SNO (CC+ES+NC) day-night spectrum data. 
We present results for both two and three 
generation scenarios. We find that the best fit is obtained
in the LMA region with the decay constant $\alpha$ zero {\it i.e.} for 
the no decay case. We had pointed earlier \cite{Choubey:2000an,
Bandyopadhyay:2001ct}
that the decay scenario was 
in conflict with the data on total rates because it suppresses low 
energy neutrinos more than the high energy ones. However  
decay also predicts 
distortion of the neutrino energy spectrum and presence of sterile 
components in the resultant neutrino beam, both of which are 
severely disfavored by the recent results from SNO and SK.
For the two generation case 
the bound at 99\% C.L. on $\alpha$ that one gets 
after including the SNO data is $\alpha \leq 7.55 \times 10^{-12} $eV$^2$.
This corresponds to a rest frame lifetime $\tau_{0}/m_{2} > 8.7 \times 
10^{-5} s/eV$. 
This is consistent with astrophysics and cosmology.   
The effect of decay is not important in the SMA region because 
the decaying fraction in the solar $\nu_e$ beam goes as $\sin^2\theta$ 
and this region remains disallowed.
There is no change in the allowed areas in the LOW region also 
because of 
eq. (\ref{galpha}) 
and the bounds on the coupling constant from K decays which 
restricts $\Delta m^2 \stackrel{>}{\sim} 10^{-6}$ eV$^2$ and
decay over the Sun-Earth distance does not take place in the LOW region.  
The three generation analysis gives stronger bounds on 
$\alpha$ for non-zero $\theta_{13}$.
In conclusion, 
the inclusion of the recent SNO data severely constrains 
decay of the solar neutrinos such that they are largely disfavored.
However nonzero values of $\alpha$ remain allowed at the 99\% C.L..

\begin{appendix}
\section{The statistical analysis of the global solar data}

We use the data on total rate from the Cl experiment, the 
combined rate from the Ga experiments (SAGE+GALLEX+GNO), 
the 1496 day data on the SK zenith angle energy spectrum and 
the combined SNO day-night spectrum data. We define the 
$\chi^2$ function in the ``covariance'' approach as
\be
\chi^2 = \sum_{i,j=1}^N (R_i^{\rm expt}-R_i^{\rm theory})
(\sigma_{ij}^2)^{-1}(R_j^{\rm expt}-R_j^{\rm theory})
\label{chi1}
\ee
where $N$ is the number of data points ($2+44+34=80$ in our case) and 
$(\sigma_{ij}^2)^{-1}$ is the inverse of the covariance matrix, 
containing the squares of the correlated and uncorrelated experimental 
and theoretical errors. The only correlated error between the total 
rates of Cl and Ga, the SK zenith-energy spectrum and the SNO day-night 
spectrum data is the theoretical uncertainty in the $^8B$ flux. 
However we choose to keep the $^8B$ flux normalization $f_B$ a free 
parameter in the theory, to be fixed by the neutral current contribution 
to the SNO spectrum. 
We can then block diagonalise the covariance 
matrix and write the $\chi^2$ as a sum of $\chi^2$ for the rates, 
the SK spectrum and SNO spectrum.
\be
\chi^2 = \chi_{\rm rates}^2 + \chi_{\rm SKspec}^2 + \chi_{\rm SNOspec}^2
\ee
For $\chi^2_{\rm rates}$ we use $R_{\rm Cl}^{\rm expt}=2.56\pm0.23$ SNU 
and  $R_{\rm Ga}^{\rm expt}=70.8\pm4.4$ SNU. The details of the theoretical 
errors and their correlations that we use can be found in 
\cite{Bandyopadhyay:2001aa,Goswami:2000wb}. 

For the 44 bin SK zenith angle energy spectra we use the data and 
experimental errors given in \cite{Fukuda:2002pe}. SK divides it's 
systematic errors into ``uncorrelated'' and ``correlated'' systematic 
errors. We take the ``uncorrelated'' systematic errors to be 
uncorrelated in energy but fully correlated in zenith angle. The 
``correlated'' systematic errors, which are fully correlated in 
energy and zenith angle, include the error in the $^8B$ spectrum 
shape, the error in the energy resolution function and the error in the 
absolute energy scale. For each set of theoretical 
values for $\Delta m^2$, $\tan^2\theta$ and $\alpha$ we evaluate 
these systematic errors taking into account the relative signs 
between the different errors \cite{Fogli:2002pt}. Finally we take an 
overall extra systematic error of $2.75\%$
fully correlated in all the bins \cite{Fukuda:2002pe}.
 
Since it is not yet possible to identify the ES, CC and NC events 
separately in SNO, 
the SNO collaboration have made available their results as a combined 
CC+ES+NC data in 17 day and 17 night energy bins.
For the null oscillation case they do give the CC and NC (and ES) rates 
\cite{Ahmad:2002jz} but firstly these rates would slightly change 
with the distortion of the $^8B$ neutrino spectrum from the Sun and 
secondly they are highly correlated. They work very well 
for studying theories with little or no energy distortion such as the 
LMA and LOW MSW solutions if the (anti)correlations between the 
CC and NC rates are taken into account and 
can be used to study the impact of the NC rate of SNO 
on the oscillation parameter space \cite{Bandyopadhyay:2002xj}. 
They can also be used to give insight 
into the $^8B$ flux and it's suppression by analysing the SK and 
SNO CC and NC data in a model independent way \cite{Bandyopadhyay:2002xj}. 
However for theories like neutrino decay where we expect large spectral 
distortions they cannot be used. Hence we analyse the full day-night 
spectrum data in this paper by adding contributions from CC, ES and NC 
and comparing with the experimental results given in \cite{Ahmad:2002jz}. 
For the correlated spectrum errors and 
the construction of the covariance matrix we follow the 
method of ``forward fitting'' of 
the SNO collaboration detailed in \cite{snodata}.

\end{appendix}

\begin{figure}
\topmargin -1in
\centerline{\epsfxsize=0.9\textwidth\epsfbox{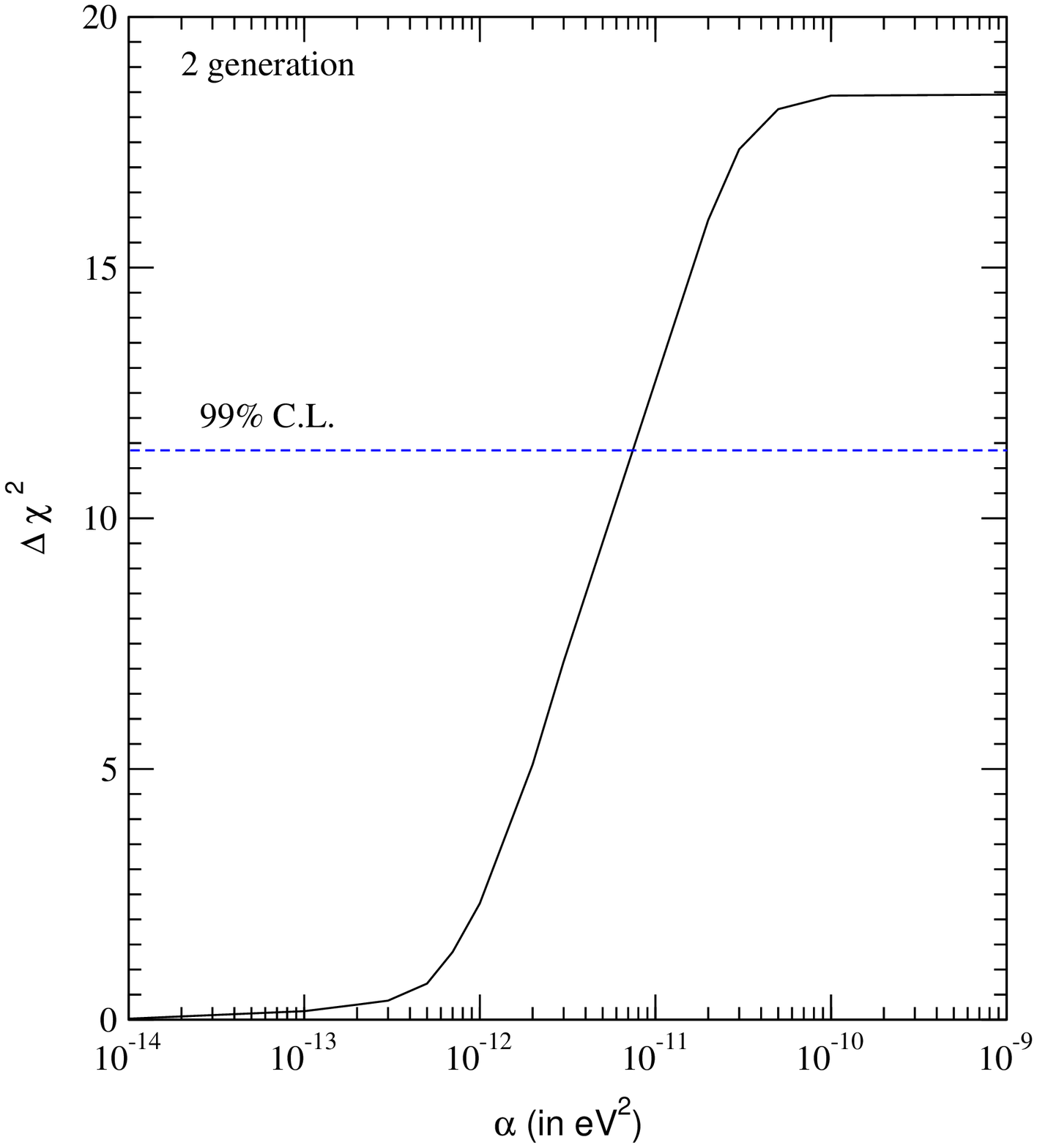}}
\vskip -0.8in
\caption{
$\Delta \chi^2 (=\chi^2 - \chi^2_{min}$) 
vs decay constant $\alpha$, for two-generations.   
Also shown is the 99\% C.L. limit 
for three-parameters. }
\end{figure}

\begin{figure}
\topmargin -2in
\centerline{\epsfxsize=0.9\textwidth\epsfbox{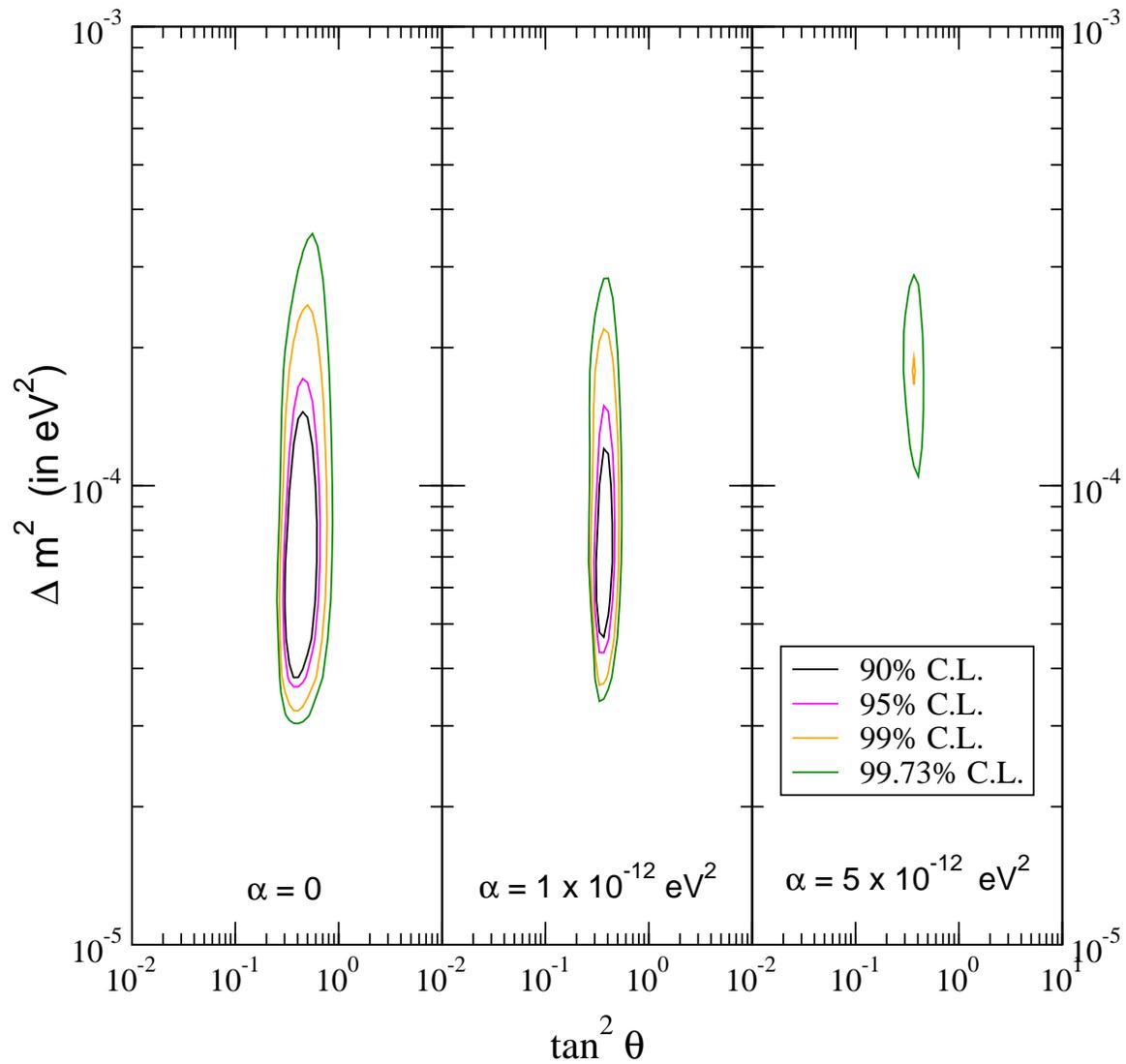}}
\vskip -1.2in
\caption{The 90, 95, 99 and 99.73\%  C.L. allowed area from the
two-generation global analysis of the total rates in Cl and Ga, 
the SK zenith angle energy spectrum data and the SNO day-night 
spectrum in presence of decay and 
oscillation. The C.L. contours are for $\Delta \chi^2$ 
corresponding to three parameters.} 
\end{figure}

\begin{figure}
\topmargin -2in
\centerline{\epsfxsize=0.8\textwidth\epsfbox{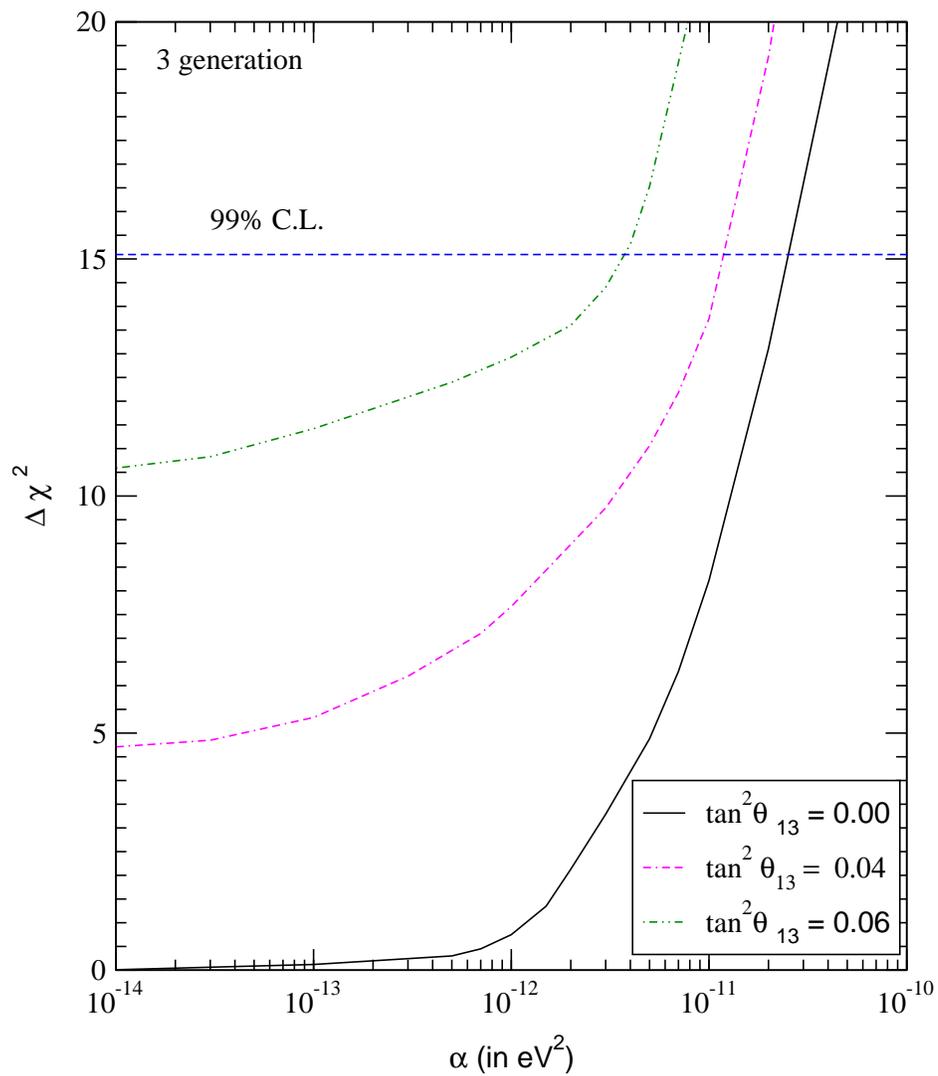}}
\vskip -0.7in
\caption{The 
$\Delta \chi^2 (=\chi^2 - \chi^2_{min}$) 
vs decay constant $\alpha$ for the three generation case 
for various fixed values of $\theta_{13}$. 
The other parameters are allowed to take any value.
}
\end{figure}

\begin{figure}
\topmargin -3in
\centerline{\epsfxsize=0.8\textwidth\epsfbox{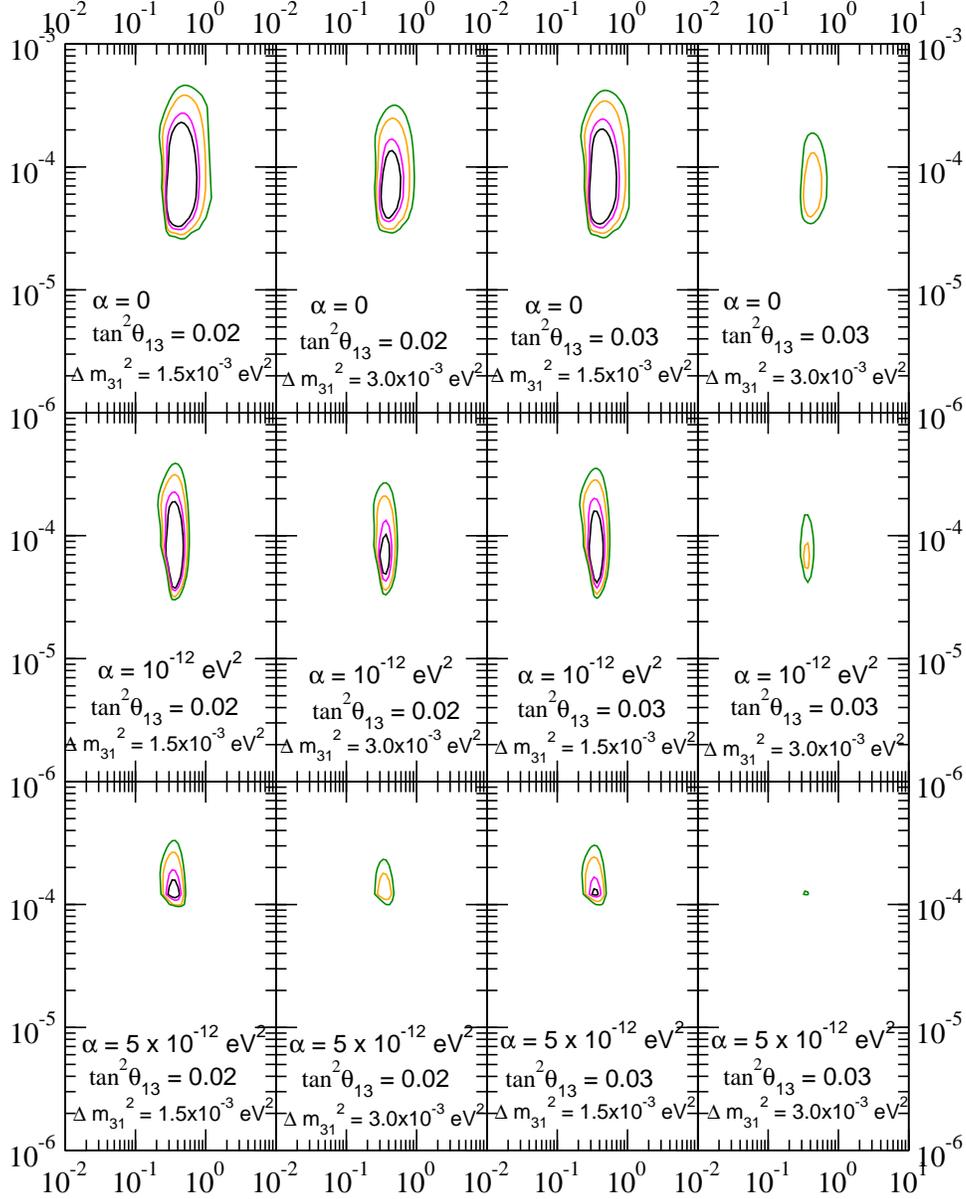}}
\vskip -0.5in
\caption{The 90, 95, 99 and 99.73\% C.L. allowed area from the
three-generation analysis of the global data 
in presence of decay and 
oscillation.}
\end{figure}

\end{document}